\documentclass[preprint,authoryear,12pt]{elsarticle}

\usepackage{graphicx}
\usepackage{epsfig}
\usepackage{amssymb}

\journal{New Astronomy}

\begin{document}

\begin{frontmatter}

\title{Magnetic field estimates for accreting neutron stars in massive binary systems and models of magnetic field decay}

\author[label1]{Chashkina, A.}
\author[label2]{Popov, S.B.}

\address[label1]{Sternberg Astronomical Institute, Moscow State University,
  Moscow, Russia 119992; Email: \emph{sagitta.minor@gmail.com}}
\address[label2]{Sternberg Astronomical Institute, Moscow State University,
  Moscow, Russia 119992; Email: \emph{polar@sai.msu.ru}}

\begin{abstract}
Some modern models of neutron star evolution predict that initially
large magnetic fields rapidly decay down to some
saturation value $\sim {\rm few}\times 10^{13}$~G and weaker magnetic fields do
not decay significantly \citep{Pons09}.
It is difficult to check the predictions of this model for initially
highly magnetized objects on the time scale of a few million years. We
propose to use Be/X-ray binaries for this purpose.
We apply several methods to estimate magnetic fields of neutron
stars in these accreting systems using the data obtained by the
RXTE satellite \citep{Galache05}. Only using the most modern approach
for estimating the magnetic field strengths of long  period NSs as proposed by \citet{Shakura11} we are able to obtain
a field distribution compatible with predictions of the theoretical
model of field decay of \citet{Pons09}.

\end{abstract}


\end{frontmatter}


\section{Introduction}

Neutron stars (NSs) are the final stage of the evolution of massive stars.  
There are different types of young isolated NSs: radio pulsars (see a review
in \citealt{ATNF}),
compact central X-ray sources in supernova remnants
\citep{Pavlov04, Luca08}, magnetars: anomalous x-ray pulsars (AXPs) and
soft gamma-ray repeaters (SGRs) \citep{Mereghetti08, Esposito11}, close-by
cooling radioquiet isolated NSs -- the
Magnificent seven (M7) \citep{Haberl07, Kaplan08}, rotating radio
transients (RRATs) \citep{Keane10}. Another large class of observed systems with NSs is formed by X-ray binaries.  They are
subdivided into high mass X-ray binaries (HMXB) and low mass X-ray binaries
(LMXB) (see a review on X-ray binaries and their evolution  in
\citet{Postnov06, Bhattacharyya10}).

In order to deeper comprehend the evolution and to follow links between different 
types of NSs we need a better understanding of the basic processes related
to NS thermal and magneto-rotational evolution.
For the initially highly magnetized NSs ($B\gtrsim \rm{few} \times 10^{13}$~G), 
modern models predict that  the
magnetic field decays significantly on the time scale of several hundred thousand years
\citep{Aguilera08a}. For magnetars there is a strong observational evidence for magnetic field decay (see, for instance, \citet{Pons07} and references therein). 
These sources are very young NSs with ages of the order
of tens of thousand years or less and with magnetic fields
$10^{14}-10^{15}$~G.  Unfortunately, it is difficult to use magnetars for testing
models of magnetic field decay on a time scale $\gtrsim 10^6$~yrs
because they become extremely elusive and avoid detection
at ages more than about a hundred thousand years.

For ``standard'' values of the initial magnetic fields ($10^{11}$ --
$10^{13}$~G) typical for radio pulsars
 the field decay in these models is insignificant: the magnetic field
diminishes just by a factor of 2. 
Therefore, the population of radio pulsars is ill-suited to test these model predictions.

\citet{Popov10} carried out a comprehensive
population synthesis of young NSs (radio pulsars, magnetars, near-by
cooling NSs) and showed that a model with magnetic field decay can
explain the basic properties of all the three populations considered using the same initial magnetic field distribution.  All
types of these objects have ages less than $\sim$ million years.  It is
important to find a way to confront model predictions and observations on a
longer time scale. NSs in some HMXBs can be suitable for testing the
evolution of magnetic fields on	 the time scales of several million years. 
Compact objects in these systems have average ages $\sim$~a few million years
(their massive companions usually have a total lifetime  less than 10-30
million years).  In the case of wind-fed accretion \citep{Negueruela10}, 
the accretion rate
is not large enough to affect significantly the evolution of the field (as
opposed to systems with intense disc accretion which produce
millisecond pulsars).  Thus, by
estimating magnetic fields of NSs in HMXBs
with accretion from a wind, it is possible to study the magnetic field
evolution on the time scale of a few million years.

The most numerous homogeneous class of HMXBs are
Be/X-ray binaries \citep{Liu06, Reig11}.  
Now there are about 70 objects of this type known in
the Galaxy, 78 in the Small Magellanic Cloud (SMC), and 16 in the Large
Magellanic Cloud \citep{Raguzova2005}.  We consider 40 Be/X-ray binaries
in the SMC to estimate magnetic fields of NSs
by means of different methods.

The article is structured as follows. 
In Sec. 2 we describe the model 
of the  magnetic field decay that we compare our estimates with. 
In Sec. 3 we present different methods to obtain estimates
of  magnetic fields of  accreting NSs. 
In Sec.  4 we
present the observational data used in the paper.  
Sec.  5 contains our results, and in Sec.  6
we discuss them.

\section{Magnetic field decay model}
In this section we describe an analytical approximation for the model of field decay used in this paper, following the approach suggested by \citet{Aguilera08a} 
and discuss types of objects that can be used for testing this model. 

There are several mechanisms for the magnetic field decay, but 
the rate of field decay is mainly determined by the Ohmic decay and the Hall drift. The
Ohmic decay involves diffusion of the magnetic field lines with respect to the charged particles. 
The characteristic Ohmic time scale is \citep{Goldreich92}:
\begin{equation}
\tau _{Ohm}=\frac{4\pi \sigma \lambda ^2}{c^2}.
\end{equation}
Here $\lambda$ is the typical magnetic field length scale and $\sigma$ is the conductivity of NS matter, $c$ is the  speed of light. 

The Hall drift time scale is:
\begin{equation}
\tau _{Hall}=\frac{4\pi e n_e \lambda ^2}{cB}.
\end{equation}
Here $n_e$ is electron density and $B$ is magnetic field strength.
The Hall drift can transport magnetic field from the inner crust where the Ohmic decay is slow, to the outer crust where it proceeds rapidly. 
Unlike the Ohmic decay, the Hall drift is essentially a non-linear process.

As we can see, the characteristic time scales of field decay
strongly depend on the length scale. This brings us to the problem of magnetic field localization. 
If the magnetic field is distributed in the entire NS, including the core, then these time
scales are long so that the magnetic field decay is insignificant on time scales about few million years. 
In the modern models  \citep{Aguilera08a, Pons09} it is assumed that the main part of the magnetic field is localized in the crust of a NS. 
In this case $\tau _{Ohm}\sim 10^5-10^6$~yrs and $\tau _{Hall} \sim (10^2-10^4)(B_{0}/10^{15}\rm{G})^{-1}$~yrs.  

The evolution of magnetic field according to the model by \citet{Aguilera08a, Pons09} is well approximated by the following equation:
\begin{equation}\label{E:Bevol}
B(t)=B_\mathrm{min}+(B_0-B_\mathrm{min}) \frac{\exp(-t/\tau_{Ohm})}{1+\tau_{Ohm}/\tau_{Hall} (1-\exp(-t/\tau_{Ohm}))}.
\end{equation}
Here  $B_0$ is the initial magnetic field strength \footnote{Below we discuss the field values at the magnetic pole:
$B\equiv B_{pole}=2B_{equator}=2\mu/R^3,$
where $R$ is the radius of a NS. For numerical estimates we use 
the value $R=10^6\rm cm$.} and $B_{min}=\rm{min}[B_0/2\ ,2 \times 10^{13}]$~G.
It can be seen that an initially strong magnetic field evolves to the asymptotic value $B=2\times 10^{13}$~G, which  can be slightly different for different objects. It is probably so in the case of magnetar SGR 0418+5729, which was born with high magnetic field strength and evolved to the value $B<7.5 \times 10^{12}$~G \citep{Esposito11}. For low-field objects the decay is not so pronounced and the field decays just by a factor of 2. 
Therefore according to the theory, there should be only a small number of objects with strong magnetic fields among populations of sources with ages $\sim 10^6$~yrs, for example among HMXBs. 

\begin{figure}
\includegraphics[width=200pt,angle=0]{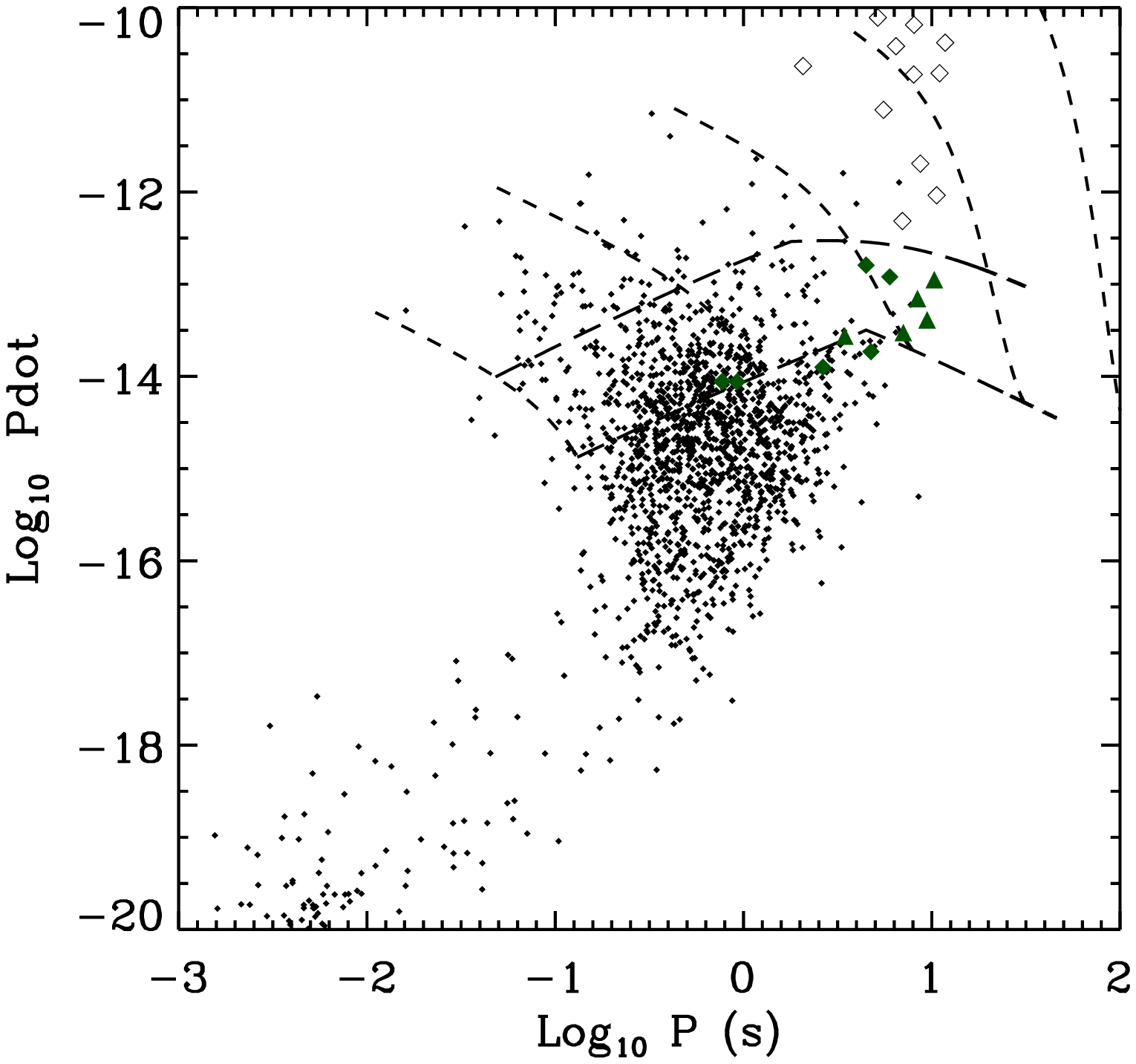}
\includegraphics[width=200pt,angle=0]{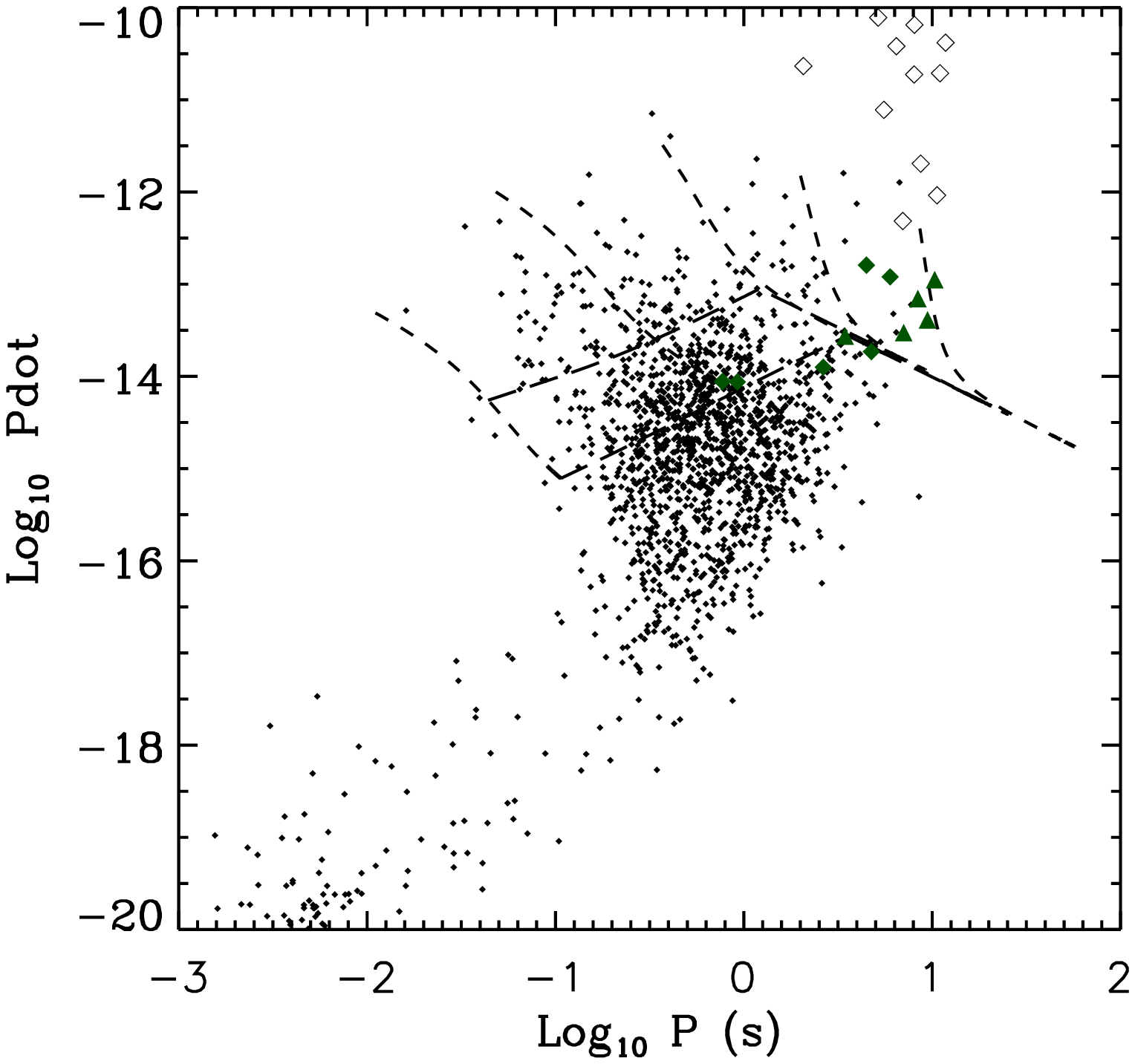}
\includegraphics[width=200pt,angle=0]{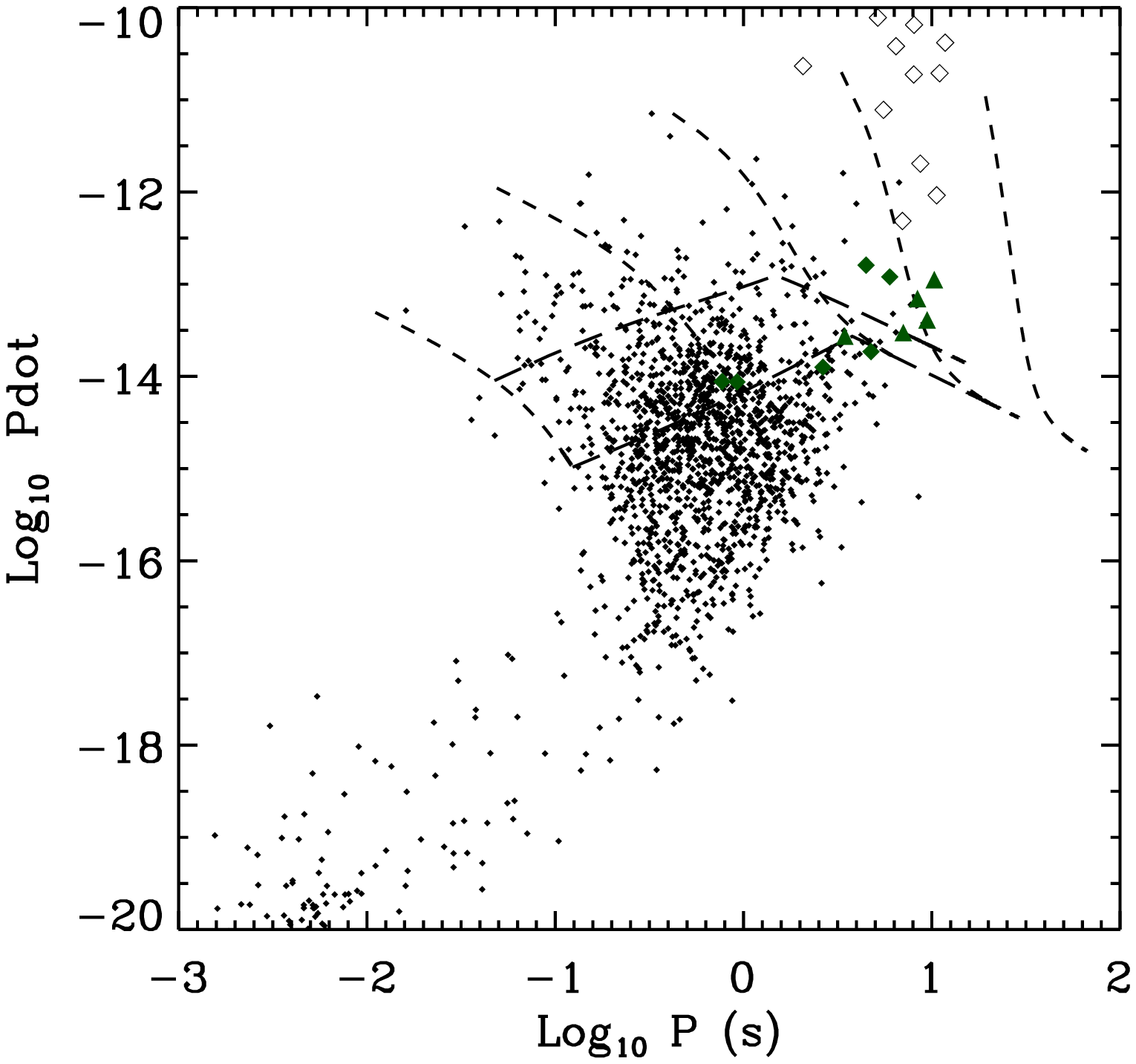}
\includegraphics[width=200pt,angle=0]{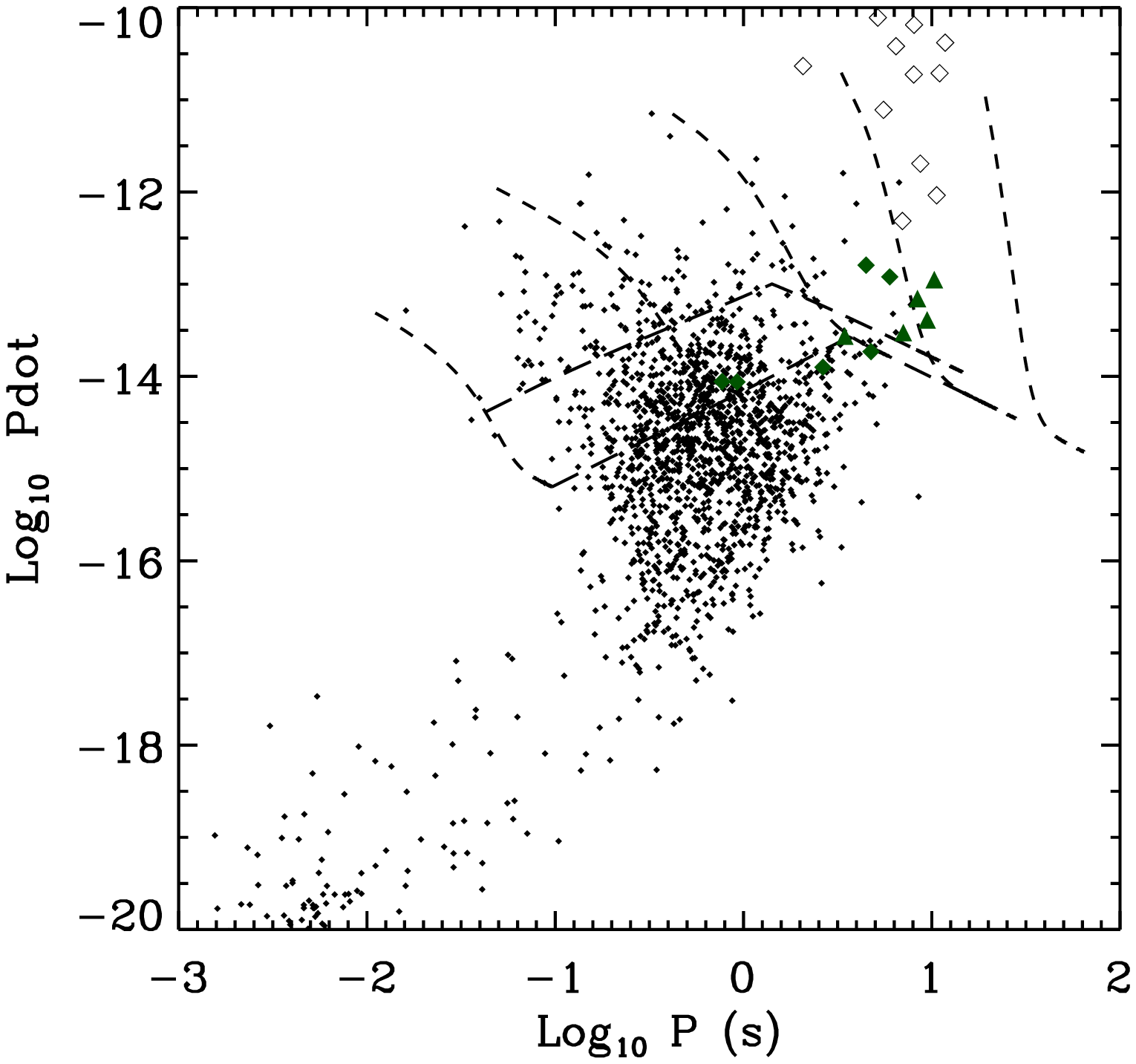}
  \caption{$P$~--~$\dot P$  diagram of radio pulsars, magnetars and radio-quiet isolated NSs. The observed populations indicated in the plots are the following: 
pulsars (black dots), magnetars (open diamonds), five of the M7 (triangles), and RRATs (filled diamonds).  In each panel NSs evolutionary tracks are shown according to analytical approximation of the field decay model  (Eq. 3) with  different initial magnetic fields from $10^{12}$~G to $10^{16}$~G. In different panels different $\tau_{Ohm}$ and $\tau_{Hall}$  are used. 
Top left: $\tau_{Ohm}=10^6$~yrs, $\tau_{Hall}=10^4(B_{0}/10^{15}\rm{G})^{-1}$~yrs.
Top right: $\tau_{Ohm}=10^7$~yrs, $\tau_{Hall}=10^2(B_{0}/10^{15}\rm{G})^{-1}$~yrs.
Bottom left: $\tau_{Ohm}=10^6$~yrs, $\tau_{Hall}=10^3(B_{0}/10^{15}\rm{G})^{-1}$~yrs.
Bottom right: $\tau_{Ohm}=10^5$~yrs, $\tau_{Hall}=10^3(B_{0}/10^{15}\rm{G})^{-1}$~yrs.
Tracks are plotted for $B_0=10^{12}, 10^{13}, 10^{14}, 10^{15}$, and $10^{16}$~G. 
In all cases NSs are treated as orthogonal rotators according to the magneto-dipole formula. 
Long-dashed lines correspond to ages $10^5$ and $10^6$~yrs.} 
\label{ppdot_analyt} 
\end{figure}

It is convenient to illustrate the evolution of the magnetic field on the $P-\dot P$ diagram.  
The results of evolution according to the eq.~(\ref{E:Bevol}) for different $\tau_{Ohm}$ and $\tau_{Hall}$ are shown in four panels of Fig.\ref{ppdot_analyt}.
 In each panel evolutionary tracks are shown for different initial magnetic fields from $10^{12}$~G to $10^{16}$~G. 
The observed populations indicated in the plots are the following: 
pulsars (black dots), magnetars (open diamonds), five of the M7 (triangles), and RRATs (filled diamonds). According to the simple model of  field decay (eq. \ref{E:Bevol}) 
sooner or later magnetic fields of initially strongly magnetized NSs converge to the asymptotic value $B\sim 2\times 10^{13}$~G, and clustering of sources can be visible. 
For PSRs, M7, or any other population this effect is not visible in the data. 
Therefore we can assume that this convergence should happen at larger ages, where no isolated NSs are observed now.

The plot for a more realistic set of parameters is shown in the top left panel: $\tau _{Ohm}=10^6$~yrs and $\tau _{Hall}=10^4(B_{0}/10^{15}\rm{G})^{-1}$~yrs. 
Therefore, we use these parameters for theoretical estimates in  sections 5 and 6. 
At the other panels the convergence is too rapid to be consistent with the observational picture.

The convergence must occur beyond the death line. Then PSRs cannot be used to test this effect.
The M7-like sources are too cold when the convergence appears. Magnetars are not active any more when the field reaches to the asymptotic value. 
We propose that determination of the magnetic fields of NSs in HMXBs can be used to see the effect of convergence on the time scale of few million years. 
We use  Be/X-ray binaries to test predictions of  theoretical models of field decay.

\section {Methods of magnetic field evaluation} 

In this section we describe 
several methods to estimate the magnetic field of accreting NSs used in this
paper.  At first, we discuss the equilibrium period hypothesis
in the case of  wind  and  disc accretion. 
Then, we formulate the models based on the values of maximal observed rates of
spin-down and spin-up of an accreting NS.  
After that we present the model based on the assumption of the equality
of Alfven and corotation radii at the moment of appearence of
pulsations.
Finally, we summarize three more sophisticated models: the model by \citet{Illarionov90}, the model by \citet{BK}, 
and a new approach based on the model by \citet{Shakura11}.
  
\subsection{The equilibrium period hypothesis}

In an accreting system a NS is affected by accelerating
and
decelerating torques (see, for instance, \citet{Lipunov92} to which we refer
for details of accretion physics basics).
The spin-up of a NS is due to the orbital angular momentum brought by the
accreted matter.  
In the
case of disc accretion the accelerating torque can be represented as a flow
of
angular momentum transported from the last Keplerian orbit:

\begin{equation}\label{E:equaldisc}
K_{su, disc}=\dot M \sqrt{GM \epsilon R_A}.
\end{equation}

For wind accretion the following equation can be used:
\begin{equation}\label{E:equalwind}
K_{su, wind}=\dot M \eta \Omega R_G^2.
\end{equation}
Here $\dot M$ is the accretion rate, $R_A=(\mu ^2/2\dot
M \sqrt{GM})^{2/7}$ is the Alfven radius, $M$ is the NS mass,
$\Omega = {2\pi}/P_{orb}$ is its orbital frequency, and
$R_G=2GM/(v_{orb}^2+v_{wind}^2)$ is the Bondi radius, 
$v_{wind}$ is the stellar wind
velocity, $v_{orb}$ is the orbital velocity, $\epsilon, \eta$ -- some numerical 
coefficients, whose values ​​depend on the properties of the acrretion flow.

To determine the Bondi radius, $R_G$, we use the orbital velocity of a NS at periastron.  For the eccentricity we
use the value $e=0.3$ as the typical one for Be/X-ray systems \citep{Reig11}.  
We take the stellar wind velocity
$v_{wind}=v_0(r/R_*)^{3/2}$ \citep{Raguzova98}, where
$v_0$ is the wind velosity at the surface of a Be-star, $r$ is the radius of
a circumstellar outflowing disc,
 and $R_*$ is the Be-star radius.  We use the following values:
$v_0=10\, \rm{km\ s^{-1}}$, $r/R_*=10$.

 The decelerating torque arises due to the magnetic field interaction with
the accretion flow.  
This torque at the stage of accretion can be estimated as follows \citep{Lipunov92}:

\begin{equation}
K_{sd}=-k_t\frac{\mu ^ 2}{R^3_{co}}.
\end{equation}
Here $\mu$ is the magnetic moment of a NS and
~$R_{co}=(GM P^2/4\pi^2)^{1/3}$ is the corotation radius, $P$ -- the spin period of NS. $k_t$ is a numerical coefficient dependent on the model of interaction between the magnetoshere and surrounding matter.

The rate of changes of the total angular momentum of
a NS can be written down as follows: 
\begin{equation} \label{E:var}
\frac{dI\omega}{dt}=\left\{
\begin{array}{rl}
\dot M \sqrt{GM \epsilon R_A}-k_t\mu ^ 2/R^3_{co}, & \mbox{ for\ \ disc\  accretion}\\
\dot{M} \eta \Omega R_G^2-k_t\mu ^ 2/R^3_{co}, & \mbox{for\ \ wind\  accretion}\\
\end{array}
\right.
\label{spinupanddown}
\end{equation}
where $I=10^{45}$~g~cm$^3$ 
 is the NS moment of inertia and $\omega=2\pi/P$ - is the spin frequency.

The equilibrium hypothesis states that an accreting NS spends most of the time in the
state with decelerating and accelerating moments nearly equalized.
When accretion starts the system rapidly evolves towards the equilibrium state (see also section \ref{sec:5}).
This state is characterized by a characteristic spin period, $P_{eq}$.
To calculate it we use the following values of numerical 
coefficients \citep{Lipunov92}:
$\epsilon = 0.45$, $\eta = 1/4$, $k_t=1/3$.
For the case of disc accretion the equilibrium period then is the following:
\begin{equation}
P_{eq.disc}=2^{15/14}\pi k_t^{1/2} \epsilon^{-1/4} \mu^{6/7} {\dot M}^{-3/7} (GM)^{-5/7}=5.47 {\mu_{30}}^{6/7} {\dot M_{16}}^{-3/7}  \mbox{s}. 
\end{equation}
For wind accretion: 
\begin{equation}
P_{eq.wind}=\sqrt{\frac{k_t\pi}{2\eta}} P_{orb}^{1/2}v^{2}(GM)^{-3/2}{\dot M}^{-1/2}\mu=47.29(P_{orb}/10\, {\rm days})^{1/2} \mu_{30}v^{2}_{8}{\dot M_{16}}^{-1/2} \mbox{s}. 
\end{equation} 
Here $\dot M_{16}$ is the accretion rate normalized to a typical value ${10^{16}}{\rm{g\, s^{-1}}}$, 
$\mu _{30}$ is the
magnetic moment normalized to a typical value $10^{30}\ \rm G\ cm^3$, and $v_8=v/10^8 \rm{cm\, s^{-1}}$ is the velocity of the accreting matter.  

Assuming that the spin period of a NS is equal to its
equilibrium period, the magnetic field $B$
for disc accretion can be estimated as follows:
\begin{equation}
B=2^{-1/4} \pi^{-7/6} k_t^{-7/12} \epsilon^{7/24} P^{7/6}{\dot M}^{1/2}(GM)^{5/6} R^{-3}.
\end{equation}  
For wind accretion:
\begin{equation}
B=2\sqrt{\frac{2\eta}{k_t \pi}} P^{-1/2}_{orb}v^{-2}(GM)^{3/2}{\dot M}^{1/2} P R^{-3}.
\end{equation} 


\subsection {Estimate based on the maximum spin-down rate}

This method is described, for instance, in \citet{Lipunov92}.

Observations show that NSs in binaries exhibit episodes of
deceleration and acceleration.  At the moment of the maximum spin-down the
decelerating torque is much larger than the accelerating one.  
In this case we can neglect the first term in eq. (\ref{E:var}) and obtain the following:
\begin{equation} \label{E:spindown}
\frac{dI\omega}{dt}=-k_t\frac{\mu ^2}{R^3_{co}}.
\end{equation}

Using the observed values of the maximum spin-down rate 
the magnetic field of a NS can be estimated as follows:
\begin{equation}
B=\frac{2}{R^3}\left(\frac{I \dot P GM}{2\pi k_t}\right)^{1/2}.
\end{equation}
This estimate should be normally considered as a lower limit, 
since we cannot be sure that no accelerating torque
exists at that moment.

The data to distinguish episodes of the maximum spin-down 
and to determine the value of $\dot P$ 
is taken from graphs in \cite{Galache05}. 
As sources are transient and accretion occurs mainly close to the periastron
passage, we use only data points
from the same episode of accretion to determine the maximum spin-down rate.

\subsection{Estimate based on the maximum spin-up rate}

The description of this approach also can be found in \cite{Lipunov92}.

At the moment of the maximum spin-up the accelerating torque is much larger than
the
decelerating one.  In this case we can neglect the second term in eq.
(\ref{E:var}) and obtain the following:

\begin{equation}
\frac{dI\omega}{dt}=\dot M (GM \epsilon R_A)^{1/2}.
\end{equation}
This equation is written for the case of disc accretion.
Using the observed values of the maximum spin-up rate, the magnetic field of NS can be estimated as follows:

\begin{equation}
B=\frac{2^{4}\pi^{7/2}}{\epsilon^{7/4}} \frac{(I\dot P)^{7/2}}{R^{3}P^{7}\dot M^{3}(GM)^{3/2}}.
\end{equation}

This estimate should be normally considered as a lower limit to the value of
the magnetic field, since we cannot be sure that no decelerating torque
exists at that moment.

We use graphs in \cite{Galache05} to find episodes of the maximum spin-up 
and to determine the value of $\dot P$ as described in the previous subsection.


\subsection{Onset of accretion and the condition $R_A=R_{co}$}

In our work we consider only Be/X-ray binaries. Many systems of this type 
have large eccentricities.  
During its orbital motion a NS can switch from the propeller to the accretor phase. At this moment the condition $R_A=R_{co}$ is
roughly satisfied:

\begin{equation}\label{E:equal}
\left(\frac{\mu^2}{2\dot M \sqrt{GM}}\right)^{2/7}=\left(\frac{GM P ^2}{4 \pi^2}\right)^{1/3}. 
\end{equation}

 We assume that at this moment pulsations start being registered.  Using
spin period and luminosity  we can estimate the magnetic field
using the eq. (\ref{E:equal}) as follows:

\begin{equation}
B = 2^{-4/7}\pi ^{-7/6} R^{-3} (\dot M)^{1/2}(GM)^{5/6}P^{7/6}.
\end{equation}

\subsection {The model by Illarionov and Kompaneets}

Above we described the well-known approaches developed by different authors in
70s-80s (see a review in \citet{Lipunov92}). Now we come to more
sophisticated models, starting with the model proposed by \citet{Illarionov90}.

This is a model with a more efficient spin-down mechanism of an accreting NS.  
The spin-down is a result of an efficient angular momentum transfer from
the rotating magnetosphere of an accreting object to the outflowing stream of
magnetized matter.  The outflow is formed within a limited solid angle, and
the outflow mass rate is less than the accretion rate. 
Frozen magnetic fields in
the outflow may provide additional angular momentum losses.

We assume, that at the moment of the 
maximum spin-down the decelerating torque
is much larger than the accelerating one.  This decelerating torque is
estimated according to the model by Illarionov and Kompaneets.  

The angular momentum losses in a solid angle $\chi$  of an outflow
 can be written as follows \citep{Illarionov90}:

\begin{equation}\label{E:illarionov}
\frac{dI \omega}{dt}=-k\frac{\chi}{2\pi} \dot M \omega R_{A}.
\end{equation}
We use this approach to obtain the decelerating torque in a small solid angle $\chi\sim 1$. For the rest of the NS magnetosphere we assume that the spin-down mechanism described above (eq.  \ref{E:spindown}) is valid.  
Then the equation for the spin evolution of a NS can be written in the
following way, neglecting the accelerating moment:
\begin{equation}
 \frac{dI \omega}{dt}=-k\frac{\chi}{2\pi} \dot M \omega R^{2}_{A}-k_t\frac{(4 \pi -\chi)\mu^2}{4\pi R^3_{co}}.
\end{equation}
To our knowledge, there is no analytical solution for this equation for magnetic field strength. Therefore we solve it numerically using bissection method.
For numerical estimates below
we use the following parameter values: $\chi = 1$, $k=2/3$, and $k_t=1/3$.
We estimate a NS magnetic field from this equation using measured $P$ and $\dot P$ which are 
 taken from graphs in \cite{Galache05}.

\subsection{The model by Bisnovatyi-Kogan}

In the model by \citet{BK} the author proposed to evaluate spin periods of NSs as:
$P=\sqrt{P_1 P_2}$. Here $P_1$ is the equilibrium period of a NS. The decelerating torque calculated according to the model by Illarionov and Kompaneets (eq. \ref{E:illarionov}) is compensated by an accelerating torque (see eq. \ref{E:equalwind}):
\begin{equation}
k\frac{\pi R_G^2}{P_{orb}}=\frac{\pi^2 R_A^3}{P^2_1}\sqrt{\frac{R_A}{2GM}}
\end{equation}
Here we use the Alfven radius in the form $R_{A}=(B^2R^6/(8\dot M\sqrt{GM}))^{2/7}$, where $B$ is the field at the magnetic pole. The numerical coefficient $k$ is assumed to be $2/3$.

The equilibrium period in this case can be calculated as follows:

\begin{equation}
P_{1}=\displaystyle\frac{\sqrt{\pi}P^{1/2}_{orb}v^2BR^3}{2^{11/4}\sqrt{k}(\dot M)^{1/2}(GM)^{3/2}}
\end{equation}

Another characteristic period, $P_2$, is  estimated in assumption that the angular velocity of the NS is equal to the angular velocity of the infalling matter on the distance of half Alfven radius:

\begin{equation}\label{E:wrong}
\frac{2\pi}{P_2}=\frac{8\pi R_G^2}{P_{orb}R_{A}^2}
\end{equation}
This equation differs from the equation given in \citep{BK} by a factor of two.

From equation (\ref{E:wrong}) the period $P_2$ can be estimated as: 
\begin{equation}
P_2=\displaystyle\frac{v^4 P_{orb}B^{8/7} R^{24/7}}{8^{4/7} 16{\dot M}^{4/7} (GM)^{16/7}}.
\end{equation}

Therefore, the spin period of a NS can be calculated as follows: 
\begin{equation}\label{E:bisn}
P=\sqrt{P_1 P_2}=0.37 \displaystyle\frac{v^3 P^{3/4}_{orb}B^{15/14}R^{45/14}}{{\dot M}^{15/28} (GM)^{53/28}}.
\end{equation}

Using relation (\ref{E:bisn}) we can estimate the magneеic field as follows:

\begin{equation}
B=2.95 \displaystyle\frac{P^{14/15}{\dot M}^{1/2} (GM)^{53/30}}{v^{14/5} P^{7/10}_{orb} R^{3}}.
\end{equation}

\subsection {Hybrid model}

Recently, a new detailed model for the case of quasi-spherical accretion was
developed by \citet{Shakura11}. 
In this model a settling subsonic accretion proceeds through a hot shell
formed around a NS magnetosphere.  The turbulent stresses are
capable of carrying the angular momentum outwards through the shell.

In this way it is possible to estimate the magnetic field
of a NS as follows:

\begin{equation} \label{E:shakura}
B_{12}=0.24\left(\frac{\delta^2\varpi \xi^{-12/33}}{1-z/Z}\right)^{11/12}\left(\frac{P/100 \rm{s}}{P_{orb}/10 \rm{days}}\right)^{11/12}{\dot M_{16}}^{1/3}{v_8}^{-11/3}.
\end{equation} 

Here we assume that the spin period is equal to the equilibrium value and $B_{12}=(B/10^{12})$~G.
 The expression in the brackets $(\delta^2\varpi \xi^{-12/33})/(1-z/Z)$ is approximately equal to 1 \citep{Shakura11}.  $\delta, \varpi, \xi, z$, and $Z$ -- are numerical coefficients.

The model developed by \citet{Shakura11}  is applicable to long period NSs with quasi-spherical accretion. Accretion in Be/X-ray binaries can proceed with or without formation of an
accretion disc around a compact object. 
It is important to distinguish objects where an accretion disc can be formed.
There is a well-known condition for accretion disc formation 
(see, for instance, \citet{Illarionov75}):
\begin{equation}
\eta \Omega R_G^2 > (GMR_A)^{1/2}.
\end{equation} 
The values of numerical coefficient $\eta$ ​​depend on the structure of stellar wind, presence of inhomogeneities and other properties of accreting matter. 
For large values of $\eta \sim 1$ an accretion disc is formed for all
of  the systems under discussion. \citet{Davies80} showed that the coefficient can be much smaller than 1. In this case an accretion disc may fail to form for  systems with large values of spin period \citep{Davies80}.  
We take into account this condition for the model by Shakura et al.  
We apply this model only to objects with $P>50$~s. 
For short-period objects ($P<50$~s) we make calculations with the 
hypothesis of the equilibrium period using standard formulae for disc accretion. So, we call this model "the hybrid".

\section{Observational data}

For our estimates we use data \citep{Galache05} on 40 Be/X-ray binaries in the 
SMC observed by the RXTE satellite. 
These data include: the spin period and count rate
(in $\rm counts\  PCU^{-1} s^{-1}$) for the pulsating component of emission. 
We convert counts to flux using 
WebPIMMS\footnote{http://heasarc.gsfc.nasa.gov/Tools/w3pimms.html}.  
Besides the standard RXTE PCA parameters we need to specify 
the Galactic hydrogen column density ($n_{H}$) towards the SMC and spectral parameters.

For the Galactic hydrogen column
density in the direction of the SMC we take the value
$n_{H}=4\cdot10^{22} \rm{cm^{-2}}$
\citep{Kuulkers07}. 
 Our results are not very sensitive to this parameter.
Even a significant variation of $n_H$ by a factor of a few results in the
magnetic field variation on the level of a few percent.

We assume that spectra can be well approximated by a single power-law. So, the only spectral parameter is the photon index. 
The photon index of all sources is taken to be equal to 1 \citep{Lipunov92}.  
The magnetic field estimates show only modest sensitivity to this parameter.
When we change a photon index by a factor of 1.5, the magnetic field changes by $\sim
10\%$.

For our magnetic field estimates we need to know the value of the accretion
rate, $\dot M$. We derive it from the X-ray flux value. Formally, we have to use
the bolometric flux, however we use a simplified approach. From the
measured count rate we derive  the flux in some output energy
range using WebPIMMS, assuming a single power-law spectrum.
To estimate the output energy
range we consider the data for different Be/X-ray systems in the Galaxy from
the paper \citep{Lutovinov08}.  In this work it is shown that most of the
objects have power-law spectra up to 25 keV, and we accept this value for
our study.
 The lower limit -- 3~keV -- for the output energy is taken from \citep{Galache05}. 
So, we have the output energy range 3-25 keV to estimate the luminosity used in
$\dot M$ calculations.  

In \citep{Galache05} only the data on pulsed flux are given. For our
estimates we need to make assumptions about the total flux, as finally we
need to derive the value of the mass accretion rate. Obviously, the ratio
between pulsed and total fluxes may be different for different sources.
However, after a statistical study we decide to use a single value for the
whole set of sources, as no details are available for most of the objects in the
sample. We assume that the total flux is 5 times larger than the pulsed flux
(Lutovinov, private communication). 

\section{Results} \label{sec:5}

\begin{figure}
\includegraphics[width=400pt]{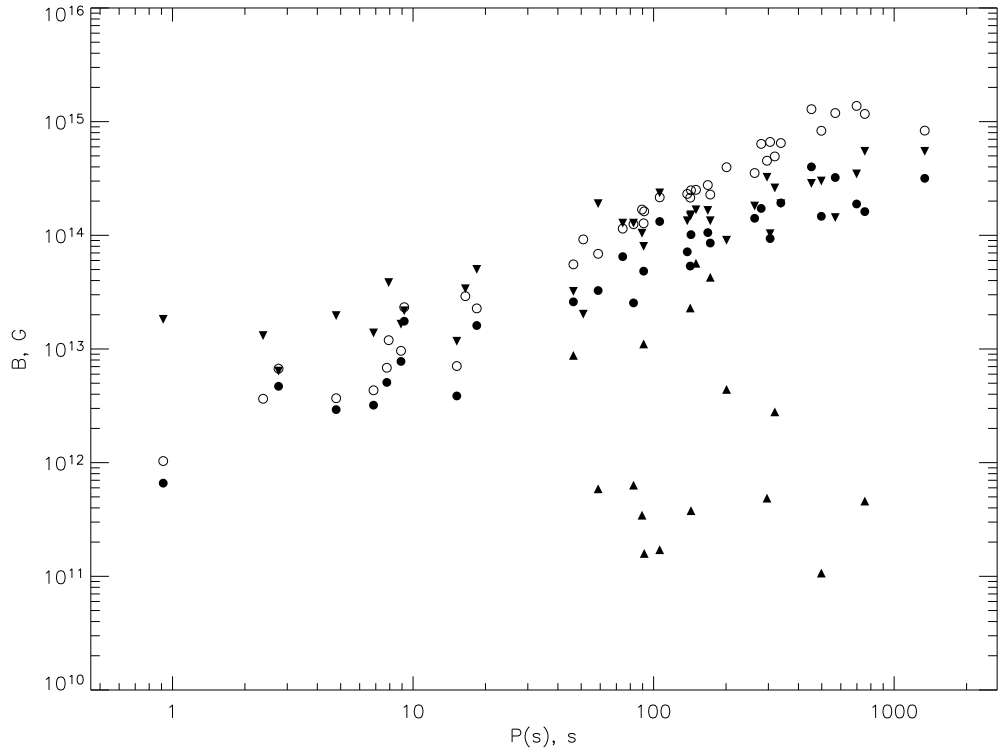}
\caption{Magnetic field values estimated using different methods: 
filled circles -- the equilibrium period hypothesis in the case of wind accretion, 
open circles -- the equilibrium period hypothesis in the case of disc accretion, 
down triangles -- the maximum spin-down, up triangles -- the maximum spin-up.}
\label{fig:M1} 
\end{figure}

\begin{figure}
\includegraphics[width=400pt]{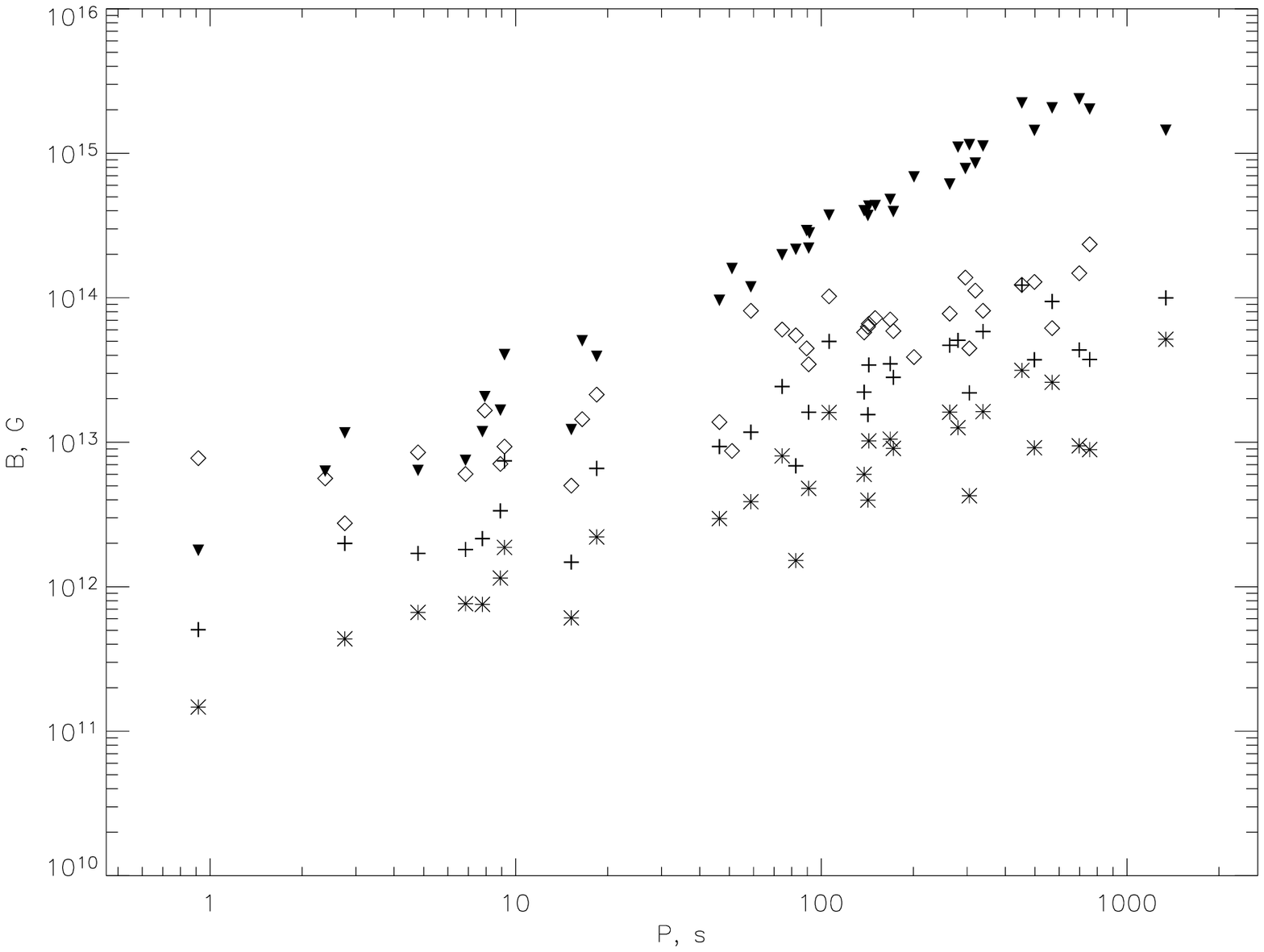}
\caption{Magnetic field values estimated using different methods (continued): 
filled triangles -- onset of pulsations when $R_A=R_{co}$,
 asterisks -- the hybrid model, diamonds -- the model by Illarionov
and Kompaneets, and crosses -- the model by Bisnovatyi-Kogan.}
\label{fig:Magnetic_fields2} 
\end{figure}

Our main results consist of estimates of NSs magnetic field in Be/X-ray binaries
obtained by several different methods described in Sec. 3. 
They are presented in  Figs. \ref{fig:M1} and \ref{fig:Magnetic_fields2}.  
Estimates for different models are shown with different symbols.
In Fig. \ref{fig:M1} estimates of magnetic field with the assumption of the equilibrium period hypothesis are plotted as filled circles, and as open circles in cases of wind and disc accretion, respectively. Estimates for the hypothesis of the  maximum spin-down and spin-up are drawn as  down triangles and up triangles, correspondingly. 
In Fig. \ref{fig:Magnetic_fields2} with filled triangles we show estimates of magnetic fields with the assumption of the onset of pulsations when $R_A=R_{co}$, and for the model by Illarionov and Kompaneets --- as diamonds. Finally, for the model by Bisnovatyi-Kogan and the hybrid model our results are given as crosses and asterisks correspondingly.

It is expected that the spin periods are on average larger for NSs with larger
magnetic fields.  This is due to the fact that for a given spin period for larger magnetic field the
decelerating torque is much larger, see eq. (\ref{E:var}), but the accelerating one remains
unchanged (for quasi-spherical accretion),
or just slightly increases (if an accretion disc is formed).  So, the
equilibrium is reached at longer spin periods for larger fields. 
The accelerating moment linearly depends on the accretion rate; in the case
of wind accretion without a disc it also depends on the properties of the
binary. This normally should result in a scatter in the relation between
spin periods and magnetic fields. 

In Figs. \ref{fig:M1}  and \ref{fig:Magnetic_fields2} we see that for
some models the correlation between spin period and magnetic field of NS is very tight.  This can be a result of a
selection effect.  In our case
pulsations are registered only above some flux value. All sources are nearly
at the same distance. Then we immediately obtain that in our approach there
is a critical value of $\dot M$, below which pulsations are undetectable. This effect is especially important for the model of equilibrium period, for the hybrid model, and for the onset of accretion when $R_A=R_{co}$.
Therefore, this effect makes the correlation $B-P$ very tight reducing the
scatter.

There is a well-known problem that for long-period objects traditional methods (equilibrium period, $R_A=R_{co}$, and the maximum spin-down, see \citet{Lipunov92}) often produce too large field value estimates. This is clearly visible in Fig.\ref{fig:M1}.
On the other hand, the model of the maximum spin-up shows more realistic results. Still, this method  strongly depends on measured parameters.  Using the data given in \citet{Galache05} we cannot measure period variations and corresponding time intervals with high precision.  The spin periods are presented in the plots with large error bars, and  time intervals corresponding to episodes of the maximum spin-up are very short. Therefore  the values of $\dot P$ are obtained with significant uncertainties, so  estimates obtained with this method are not very certain. However, we present results based on the maximum spin-up for long-period sources for which estimates seem to be reliable.

Estimates based on the model by Bisnovatyi-Kogan are more reliable than those based on the maximum spin-up, and do not contain very high values $\gtrsim 10^{14}$~G. Therefore, there are many NSs with $B>3\times 10^{13}$~G in contradiction with theoretical expectations. 

Finally, the hybrid model is able to reproduce most of the predicted deficit of long-period high magnetic field objects. 
  Thus, we consider this model to be the best among studied and  discuss results based on it in the next section in more details.

\section{Discussion}

In Fig. 4 we show magnetic field distributions. Here the theoretical distribution for the initial polar magnetic field $B_0$ is generated randomly for $10^5$ NSs from a log-normal distributon with the mean value $<\log{B_0/[G]}>=13.25$ and standard deviation $\sigma_{\log{B_0}}=0.6$ \citep{Popov10}. Note, that formally this initial field distribution is not absolutely self-consistent with our model of field decay, because we use parameters slightly different from those used in numerical calculations by \citep{Popov10}. However, this imperfection cannot significantly influence our conclusions, because the main assumption is related to the asymptotic field value.

\begin{figure}
\includegraphics[width=200pt]{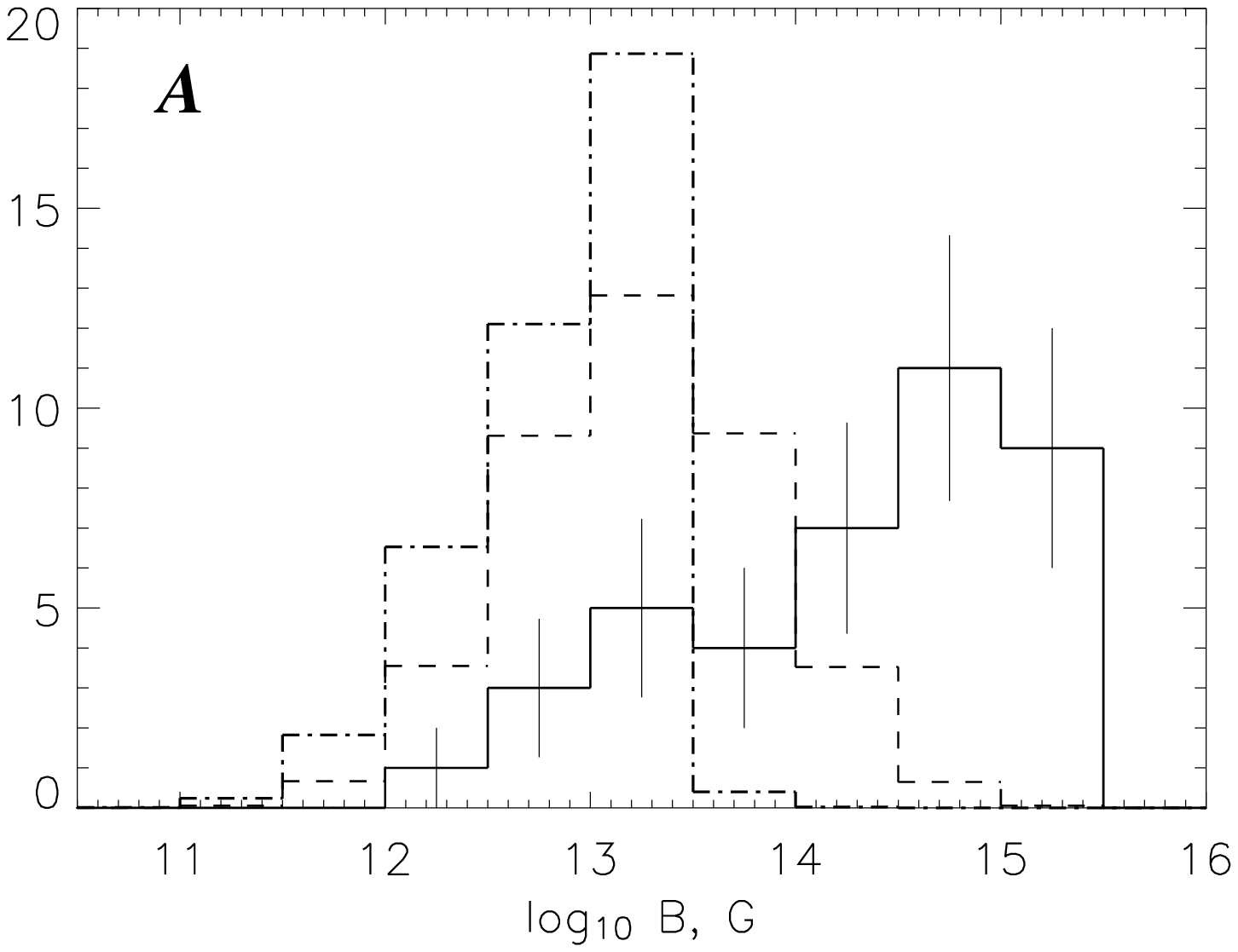}
\includegraphics[width=200pt]{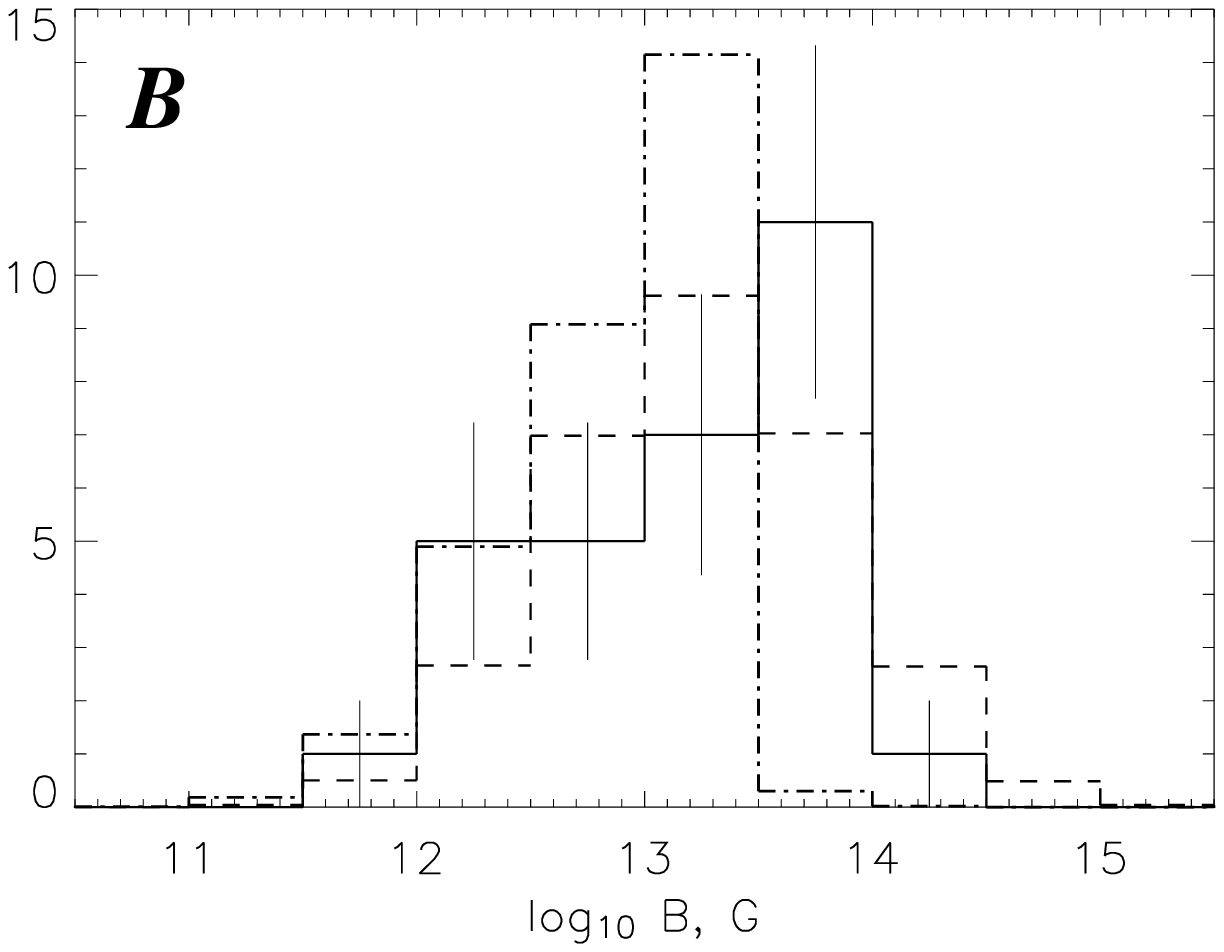}
\includegraphics[width=200pt]{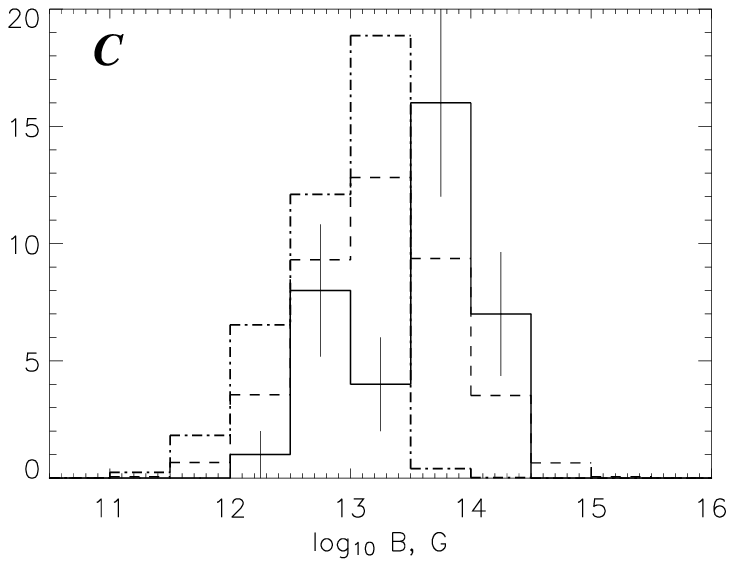}
\includegraphics[width=200pt]{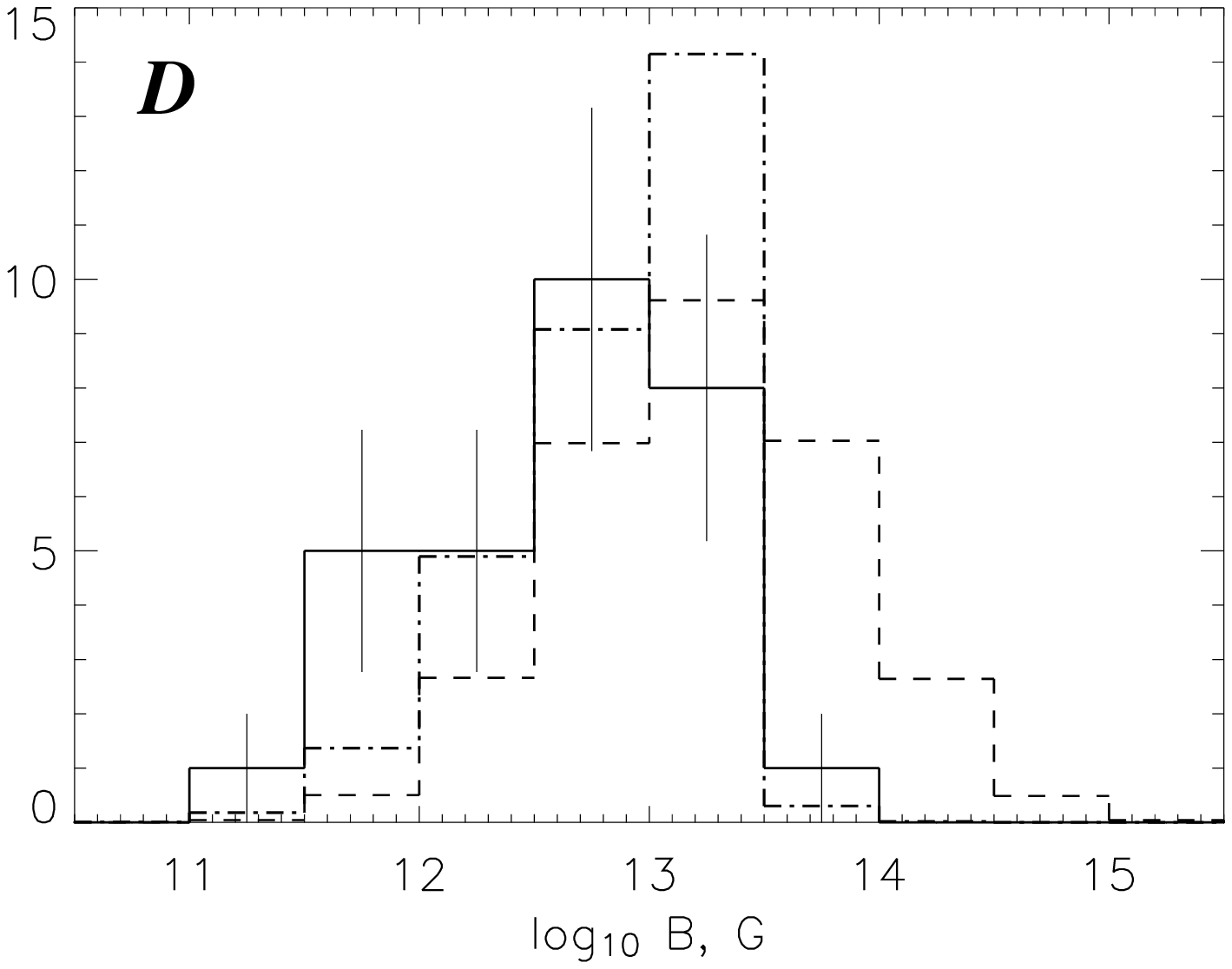}
\caption{Magnetic field distributions for different models (solid lines). Panel A -- onset of pulsations when $R_A=R_{co}$. Panel B -- the model by Bisnovatyi-Kogan. Panel C -- the model by Illarionov and Kompaneets. Panel D -- the hybrid model. Theoretical models for the initial and evolved magnetic fields are shown in all the four panels by dashed and dash-dotted lines, respectively. All distributions are normalized to the total number of observed objects. Error bars are given according to the Poissonian values.
}
\label{fig:mfe} 
\end{figure}

The evolved fields  are calculated  according to eq. (\ref{E:Bevol}) using the values: $\tau _{Hall}=10^4(B_{0}/10^{15}\rm{G})^{-1}$~yrs and $\tau _{Ohm}=10^6$~yrs. The  NS ages are chosen uniformly in the interval $[0, t_{max}]$, where $t_{max}=10^7 \rm$~yrs is the maximum life time of Be/X-ray binaries. Magnetic field estimates obtained in this paper are shown in four different panels in Fig. \ref{fig:mfe}. In the  panel A with a solid line we show the  magnetic field distribution based on the assumption of the onset of pulsations when $R_A=R_{co}$. The magnetic field distribution according to the model by Bisnovatyi-Kogan is plotted in the panel B. Distributions for the models by Illarionov, Kompaneets and the hybrid model are shown in the panels C and D, correspondingly.
Theoretical models for the initial and the evolved magnetic fields are shown in all four panels by dotted and dot-dashed lines, correspondingly.

To compare the theoretical predictions with results based on  the hybrid model we use the Kolmogorov-Smirnov test.
According to this test, the hypothesis of the same distribution of the theoretical and estimated values cannot be rejected at 80\% significance level. However,  there is a small number of objects for the hybrid model distribution in the bin $10^{13}<B<10^{13.5}$~G (Fig. \ref{fig:mfe}D). Discrepancies can be related to uncertainty in evolution of the magnetic field for $t>10^6$~yrs. Probably, the magnetic field decay continues but more slowly than before.  Likewise, different highly magnetized objects can evolve to different values of the asymptotic magnetic field (see Sec. 2).

In this study we assumed that NSs in binaries have the same initial magnetic distribution as isolated compact objects. This is not so obvious in the case of magnetars, as in the stardard scenario their magnetic fields are 
generated via the dynamo mechanism \citep{dt1992}. The dynamo mechanism is effective in the case of very rapid rotation of a protoNS. Isolated massive stars hardly can have rapidly rotating cores. Rapid rotation can be reached in close binary systems due to accretion or/and tidal synchronization. On the other hand, all known magnetars are isolated objects, and authors studied evolutionary channels in binary systems which result is formation of isolated compact objects with rapid initial rotation \citep{bp2009, pp2006}. In these models objects with extremely large field tend to form after a binary is destroyed to explain that all known magnetars are isolated. 

\begin{figure}
\includegraphics[width=350pt]{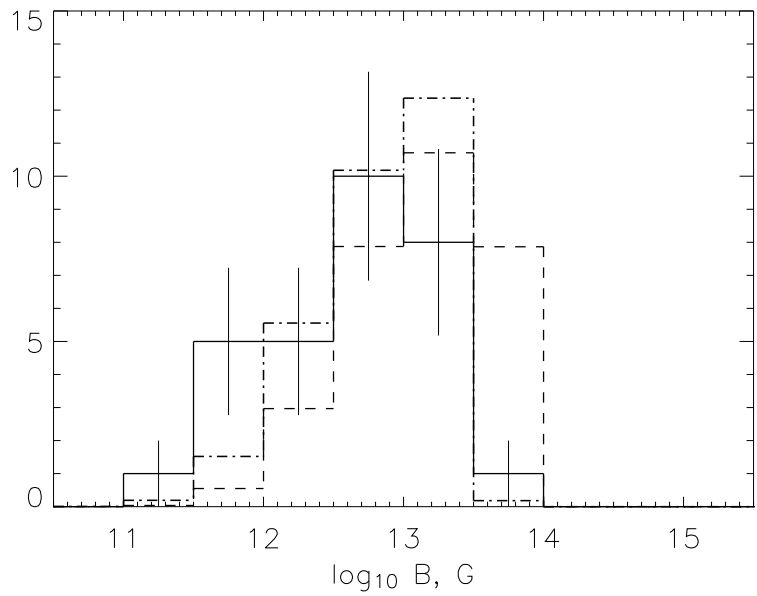}
\caption{Magnetic field distributions. The distribution based on observational
data after application of the hybrid model is shown with a solid line.
Poissonian error bars are indicated. Results for the theoretical model
for evolved magnetics fields is shown with a dash-dotted line.
Finally, a modification of the theoretical model for which a cut-off
at $10^{14}$~G for initial fields was used is shown with a dashed
histogram. All distributions are normalized to the total number of observed objects.}

\label{fig:mfe2} 
\end{figure}

To test the possibility of absence of extreme magnetars in binary systems for the case of the hybrid model (which is the most successful in explaining the properties of the population under study), we plot an additional histogram (Fig. \ref{fig:mfe2}) for a modified initial magnetic field distribution: no NSs with $B_0>10^{14}$~G are formed. As we see, estimates made on the basis of observational data are even in better agreement with the prediction for the evolved field in the model by \citet{Pons09} with the cutoff at $B_0=10^{14}$~G, as the discrepancy in the bin $10^{13}<B<10^{13.5}$~G is decreased. However, we note that our calculations are based on an approximation to the actual detailed model by \citet{Pons09}, and for the specific choice of parameters, which can vary ($\tau_{Hall},\tau_{Ohm}, B_{min}$). We do not overestimate the accuracy of our results, and so do not make further speculations.

An additional consistency check for magnetic field estimates can be made using the Corbet diagram \citep{Corbet84}. We apply this test to the hybrid model. This diagram relates spin periods and orbital periods of NSs. We simulate this distribution. Orbital periods of NSs are chosen uniformly in the interval [10, 1000] days. Using an orbital period and evolved magnetic field we can estimate the spin period of a NS. We calculate the evolved magnetic field as discussed above. These values of spin and orbital periods are plotted in Fig. 6 as dots. We plot the observational data \citep{Reig11} as diamonds in this figure. One can see good agreement between theoretical and observational data. According to the non-parametric Kolmogorov-Smirnov test, the hypothesis that both subsets arise from the same distribution cannot be rejected at 77\% significance level.

The methods that we use to estimate  magnetic fields have limitations, especially the
most simple ones. For instance, the hypothesis of the 
equilibrium period may not be valid for some systems.  In our
estimates we use $\dot M$ obtained for the moment
when pulsations are detected for the first time.
But for NSs in 
systems with high eccentricity which spend much time at the propeller
stage, it can take some time to reach the 
equilibrium. So, their observed periods are larger than the equilibrium
value. This results in underestimating of the magnetic field.

Likewise, we should consider that both disc and wind accretion estimates (see, Fig. \ref{fig:M1}) cannot be correct at the same time because just one regime is realized in any given source.
On average, estimates for disc accretion are more realistic for shorter spin periods. Still, we think that estimates based on old models are not realistic in general, as they produce too large field values for long period sources.  

In the model by \citet{Shakura11} there is a critical value of $\dot M$. For accretion rates
above this $\dot M_{cr}$ an envelope around a compact object can cool rapidly. This results in an enhanced accretion rate, and no equilibrium is possible in this regime. This critical value depends on several parameters and can be different in different systems, but roughly $\dot M_{cr}\lesssim 10^{17}$~g/s for magnetic fields $\gtrsim 10^{13}$~G.  
Marginally, this condition is fulfiled for our systems. We can slightly overestimate the accretion rate, however, if the model by \citet{Shakura11} is applicable, then the magnetic field estimate depends on the accretion rate only mildly.

 It is instructive to compare the results of indirect magnetic field estimates with direct magnetic field measurements through cyclotron lines, as it was made for the long-period object GX 301-2. For this object, magnetic field strengh is estimated by \citet{LaBarbera2005} as  $\sim 5\times 10^{12}$~G. This value is consistent  with the magnetic field estimate made by \citet{Shakura11} using their model. On the other hand, the classical models used by \citet{Doroshenko10}
provide the magnetic field estimate nearly two orders of magnitude larger
than it is given by the cyclotron line measurements.  However, the cyclotron line can be formed high above the NS surface that allows the measured magnetic field strength to be considerably weaker than that on the surface \citep{Doroshenko10}.

\begin{figure}
\includegraphics[width=400pt]{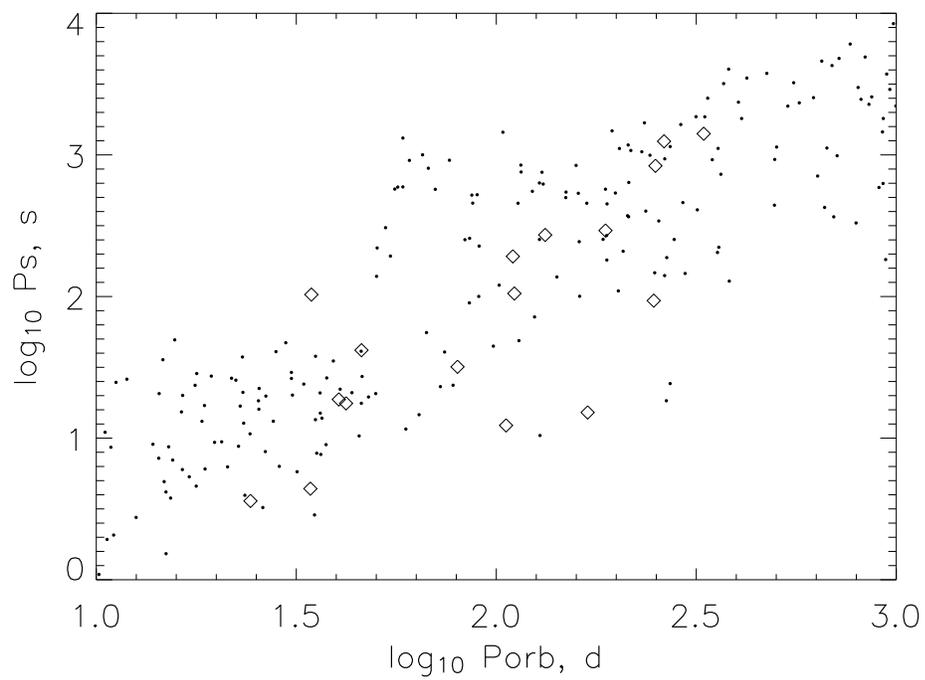}
\caption{Corbet diagram for the observed data (open diamonds) and for the hybrid model (dots).}
\label{fig:mfe3} 
\end{figure}

Finally, we comment on a possible evolution of the angle between spin and magnetic axis.  
For radio pulsars it is possible to describe the angular momentum evolution using the magneto-dipole formula \citep{Pacini67}:
\begin{equation} \label{E:angle}
\frac{dI \omega}{dt}=-\frac{1}{6}\frac{B^2\omega ^3 R^6}{c^3}\rm{sin}^2\beta,
\end{equation}
Here $\beta$ is the angle between spin and magnetic axis. From this formula one can estimate the effective magnetic field $B_{eff}=B \rm{sin}\beta$. So, the effect of the variation of the effective magnetic field can be related not only to the magnetic field decay, but also to evolution of the angle between magnetic and spin axis. Therefore, it is difficult to distinguish between these effects. However, there are different methods to estimate magnetic field of magnetars and to demonstrate that this is really the field decay, not the variation of the angle $\beta$.
Note, that our methods are only weakly sensitive to the angle between spin and magnetic axis, and we estimate the absolute value of magnetic field.  Still, using Be/X-ray systems we can try to study evolution of the angle between these axis. 
According to two most popular models (magneto-dipole and current losses), the angle evolves very fast on the time scale of spin period evolution. The angle $\beta$ evolves to the value $\beta=0^{\circ}$ in the magneto-dipole model and to the value $\beta=90^{\circ}$ in the current losses model \citep{Beskin93}.  But this effect is not observed for PSRs.
It is possible to use Be/X-ray binaries to study evolution of the angle on the time scale up to $10^7$~yrs jointly with the magnetic field decay. We plan to use this approach in future studies.

\section{Summary}
In our work we used different models to estimates magnetic fields of NSs in Be/X-ray binaries in the SMC. Most of them are shown to overestimate the magnetic field, but the hybrid
 model based on the approach by \citet{Shakura11} gives good agreement with the prediction of the theoretical model of magnetic field decay on a time scale up to $\sim 10^7$ yrs. 

\section*{Acknowledgements}
We thank N.I. Shakura and K.A. Postnov for the opportunity to use their model before publication,  and for consultations. We acknowledge A.A. Lutovinov and J.A. Pons for interesting discussions. We thank P.K. Abolmasov and E.A. Vasiliev for useful remarks. Also we are grateful to the anonymous referees for useful comments. Sergei Popov thanks University of Padova for support and hospitality during his visit.
This work was supported by the Russian Foundation for Basic Research grants 10-02-00599, 12-02-00186 and by the Federal program for research staff (02.740.11.0575).

\bibliographystyle{model2-names}
\bibliography{ref}

\end{document}